\documentclass[aps,pra,twocolumn,showpacs,superscriptaddress,tightenlines,sort&compass,amsmath,amssymb,floatfix]{revtex4}

\usepackage{comment}
\usepackage{amssymb}
\usepackage{amsmath}
\usepackage[dvipsnames,usenames]{color}
\usepackage{graphicx,graphics}
\usepackage{amssymb}

\def\extra#1{{}}
\def \etal{\textit{et al.}}

\newcommand\ket[1]{\ensuremath{|#1\rangle}}
\newcommand\bra[1]{\ensuremath{\langle#1|}}

\begin{document}

\title{Efficient tomography of quantum-optical
Gaussian processes probed with a few coherent states}

\author{Xiang-Bin Wang}
\email{xbwang@mail.tsinghua.edu.cn} \affiliation{Department of
Physics and  Key Laboratory of Atomic and Nanosciences, Ministry
of Education, Tsinghua University, Beijing 100084, China}
\affiliation{CEMS, RIKEN, Saitama, 351-0198, Japan}
\affiliation{Jinan Institute of Quantum Technology, Shandong
Academy of Information and Communication Technology, Jinan 250101,
China}
\author{Zong-Wen Yu}
\affiliation{Department of Physics and  the Key Laboratory of
Atomic and Nanosciences, Ministry of Education, Tsinghua
University, Beijing 100084, China}
\affiliation{Data Communication Science and Technology Research Institute, Beijing 100191, China}%
\author{Jia-Zhong Hu}
\affiliation{Department of Physics and  the Key Laboratory of
Atomic and Nanosciences, Ministry of Education, Tsinghua
University, Beijing 100084, China}
\author{Adam Miranowicz}
\affiliation{CEMS, RIKEN, Saitama, 351-0198, Japan}
\affiliation{Faculty of Physics, Adam Mickiewicz University,
PL-61-614 Pozna\'n, Poland}
\author {Franco Nori}
\affiliation{CEMS, RIKEN, Saitama, 351-0198, Japan}
\affiliation{Physics Department,The University of Michigan, Ann
Arbor, Michigan 48109-1040, USA }

\begin{abstract}
An arbitrary quantum-optical process (channel) can be completely
characterized by probing it with  coherent states using the
recently developed coherent-state quantum process tomography (QPT)
[Lobino {\em et al.}, Science {\bf 322}, 563 (2008)]. In general,
precise QPT is possible if an infinite set of probes is available.
Thus, realistic QPT of infinite-dimensional systems is approximate
due to a finite experimentally-feasible set of coherent states and
its related energy-cut-off approximation. We show with explicit
formulas that one can completely identify a quantum-optical
Gaussian process just with a few different coherent states without
approximations like the energy cut-off. For tomography of
multimode processes, our method exponentially reduces the number
of different test states,  compared with existing methods.
\end{abstract}


 \maketitle

\section{Introduction}

One of the basic problems of quantum physics is to predict the
evolution of a quantum system under certain conditions. For an
isolated system with a known Hamiltonian, the evolution is
characterized by a unitary operator determined by the
Schr\"odinger equation. However, the system may interact with its
environment, and the total Hamiltonian of the system plus the
environment is in general not completely known. The evolution can
then be regarded as  a ``black-box
process"~\cite{Poyatos97,Chuang97,DAriano01} which maps the input
state into an output state. An important problem here is how to
characterize an unknown process by testing the black-box with some
specific input states, which is referred to as quantum process
tomography (QPT) (for reviews see Refs.~\cite{Paris04,Mohseni08}).

QPT can be understood as the tomography of a quantum channel since
any physical operation describing the dynamics of a quantum state
can be considered as a channel~\cite{Eisert05}. In contrast, the
goal of quantum state tomography (QST) is the reconstruction of an
unknown state (i.e., its density matrix) by a series of
measurements on multiple copies of the state (for a review see
Ref.~\cite{Paris04}). Both QPT and QST are essential tools in
quantum engineering and quantum information processing.

A few methods for QPT were developed, including the standard
QPT~\cite{Chuang97,Poyatos97}, ancilla-assisted
QPT~\cite{DAriano01,Leung03,Altepeter03,DAriano03}, direct
characterization of quantum dynamics~\cite{Mohseni06,Wang07}, and
coherent-state QPT~\cite{Lobino08}. There are dozens of proposals
and experimental realizations of QPT for systems with a few
qubits. These include the estimation of quantum-optical
gates~\cite{Altepeter03,Mitchell03,Demartini03,Obrien04,
Nambu05,Langford05,Kiesel05,Wang07,Ma12}, liquid
nuclear-magnetic-resonance
gates~\cite{Childs01,Boulant03,Weinstein04}, superconducting
gates~\cite{Liu04,Zagoskin06,Neeley08,Chow09,Bialczak10,Chow12}
(for a review see Ref.~\cite{You11}) and other solid-state
gates~\cite{Kampermann05,Howard06,Burgarth11}, ion-trap
gates~\cite{Riebe06,Monz09}, or the estimation of the dynamics of
atoms in optical lattices~\cite{Myrskog05}. In contrast, there are
only a very few experimental demonstrations of QPT for
infinite-dimensional systems (see, e.g., Ref.~\cite{Lobino08}).

Any physical process can be described by a completely positive map
$\varepsilon$. Such a process is fully characterized if the
evolution of any input state $\rho_{\rm in}$ is predictable:
$\rho_{\rm out}=\varepsilon (\rho_{\rm in})$. In general, QPT is
very difficult to implement in high-dimensional spaces,  and, more
challengingly, in an infinite-dimensional space, such as  a Fock
space~\cite{DAriano03,Lobino08}. Recently, Ref.~\cite{Lobino08}
described QPT in a Fock space for continuous variable (CV) states.
Two conclusions can be drawn~\cite{Lobino08}: ($i$) If the output
states of {\em all} coherent input states are known, then one can
predict the output state of any input state; ($ii$) By taking the
photon-number-cut-off (or energy cut-off) approximation, one can
then characterize an unknown process with a finite number of
different input coherent states (CSs).

It is an interesting question to identify an exact QPT with a
finite number of coherent states. If the process is completely
unknown, then QPT with a finite number of coherent states is
impossible. However, if some of the constraints of the quantum
process are known, then QPT can be simplified and, thus,
effective. Gaussian maps are the most common for quantum-optical
processes. In this article, we show that if a certain quantum
process is known to be Gaussian, then an exact QPT can be
performed with only a few different coherent states.

It is worth noting that there is an analogy between QPT and QST,
especially for quantum-optical Gaussian processes
(channels)~\cite{Eisert05,Holevo12} and Gaussian
states~\cite{Holevo11,Rehacek09}. This analogy can be seen, e.g.,
by comparing correlations between observables encoded in the
covariance matrices, which completely describe a Gaussian object
(either a quantum state or process). Thus, tomographies of quantum
Gaussian systems are effectively finite-dimensional with their
covariance matrix having a physical meaning analogous to a
finite-dimensional density matrix.

As has been shown in Ref.~\cite{DAriano01}, QPT can be performed
with two-mode squeezed vacuum (TMSV) for any unknown process.
However, TMSVs are not so easy to manipulate in practice,
especially, because this involves quantum tomography of entangled
states, which is not an easy task.

Here, we show that based on existing results~\cite{DAriano01}, by
using the standard quantum-optical Husimi $Q$-representation, one
can perform QPT with only a few CSs without entangled ancillas for
quantum-optical Gaussian processes. The method described here has
several advantages. First, it presents explicit formulas {\em
without} any approximations, such as the photon-number-cut-off
approximation. Second, it requires only a few different states to
characterize a process, rather than {\em all} CSs. Third, for
multimode Gaussian process tomography, the number of input CSs
increases {\em polynomially} with the number of modes, rather than
exponentially. Fourth, it uses the Husimi $Q$-functions only,
which is always well-defined for {\em any} state without any
higher-order singularities in the calculation.

The paper is organized as follows: We review the existing results
about QPT \emph{based on} entangled ancillas in Sec.~II. In
Sec.~III, the QPT \emph{without} ancillas is proposed for
single-mode Gaussian processes. A simple illustrative example of
the method is discussed in Sec.~III.A. A generalization of our QPT
for a multi-mode case is presented in Sec.~IV. We conclude in Sec.
V.

\section{Ancilla-assisted quantum process tomography}

First, we review the existing result of the ancilla-assisted QPT
with TMSV~\cite{DAriano01} to show some similarities but also
crucial differences in comparison to our proposal of ancilla-free
QPT, which will be described in Sec.~III.

A TMSV is defined by $|\chi(q)\rangle=c_q \exp({q a^\dagger
b^\dagger})|00\rangle$, where $c_q=\sqrt{1-q^2}$, and $q$ is real.
The (unnormalized) maximally-entangled state here is
\begin{equation}\label{eq:MaxEntState}
  |\Phi^+\rangle =\lim_{q\rightarrow 1} \exp(qa^\dagger
  b^\dagger)|00\rangle=\sum_{k=0}^\infty |kk\rangle,
\end{equation}
where $a^\dagger$ ($b^\dagger$) is the creation operator for mode
$a$ ($b$). Note that entanglement is not required for the
ancilla-assisted QPT, but it makes it more efficient. In
particular, the use of the maximally-entangled states can make the
QPT experimentally optimal with regard to perfect nonlocal
correlations~\cite{Altepeter03}.

Assume now that the black box process acts only in mode $b$ of the
bipartite state $|\chi(q)\rangle$. After the process, we obtain a
two-mode state $\Omega_q$. One can define the projection operator
\begin{equation}\label{eq:ProOpt}
  T(q) =c_q \exp[(\ln q)a^\dagger a],
\end{equation}
which has the property~\cite{Louisell73}:
\begin{equation}\label{eq:ProOptProp}
  T(q)\;(a,a^\dagger)\;T^{-1}(q)=(a/q,qa^\dagger).
\end{equation}
The TMSV $|\chi(q)\rangle$ can be written as
\begin{equation}\label{tqlink}
\ket{\chi(q)}=T(q)\otimes I \ket{\Phi^+}.
\end{equation}
According to Eq.~(\ref{tqlink}), we have
\begin{equation}
\Omega_q = T(q) \otimes I\cdot \rho_{\varepsilon}\cdot T(q)\otimes
I,
\end{equation}
where $\rho_{\varepsilon}=I\otimes \varepsilon
\left(|\Phi^+\rangle\langle\Phi^+|\right)$. Naturally,
\begin{equation}\label{phiq}
\rho_{\varepsilon} = T^{-1}(q) \otimes I \cdot  \Omega_q \cdot
T^{-1}(q)\otimes I.
\end{equation}
We now also formulate the output state of any single-mode input
state $|\psi(\{c_k\})\rangle =\sum_k c_k |k\rangle$ of mode $b$.
Obviously it can be written as
\begin{eqnarray}\label{00}
\left(|\psi\rangle\langle\psi|\right)_b &=& {_a}\langle \psi^*|
\Phi^+\rangle\langle \Phi^+|\psi^*\rangle_a\nonumber\\
&=&{\rm tr_a}\left(|\psi^*\rangle\langle\psi^*|\otimes I\cdot
|\Phi^+\rangle\langle \Phi^+|\right),
\end{eqnarray}
and $|\psi^*\rangle_a=\sum_k c_k^*|k\rangle_a$ is a single-mode
state for mode $a$ (sometimes we omit the subscript $a$ or $b$ for
simplicity). We obtain the output state
\begin{eqnarray}\label{e12}
\rho_\psi &=& {_a\langle
\psi^{*}|\rho_{\epsilon}|\psi^{*}\rangle_a
=\rm{tr}_{a}(|\psi^{*}\rangle\langle\psi^{*}|\otimes I\cdot \rho_{\epsilon})} \nonumber \\
&=& {\rm tr_a} \left[ |\psi^*(\{c_k/q^k\})\rangle\langle
\psi^*(\{c_k/q^k\})|\otimes I \cdot \Omega_q\right] \nonumber\\
&=& {_a\langle \psi^*(\{c_k/q^k\})| \Omega_q
|\psi^*(\{c_k/q^k\})\rangle_a}.
\end{eqnarray}
More explicit expressions can be obtained by using the Husimi
$Q$-function. If the single-mode input state in mode $b$ is a
coherent state $|\alpha\rangle$,  the output state then becomes
\begin{equation}
\rho_{\alpha} =\langle \alpha^*|  \rho_{\varepsilon}
|\alpha^*\rangle= \langle \alpha^*|T^{-1}(q)\otimes I \cdot
\Omega_q \cdot T^{-1}(q)\otimes I|\alpha^*\rangle .
\end{equation}
Note that the state $|\alpha^*\rangle$ here is a single-mode
coherent state in mode $a$. Using the property of $T(q)$ and the
definition of  CSs, $a|\alpha^*\rangle =
\alpha^*|\alpha^*\rangle$, we easily find
\begin{equation}
T^{-1}(q)\otimes I\;|\alpha^*\rangle  = \mathcal N_q(\alpha)\;
|\alpha^*/q\rangle,
\end{equation}
where the factor $\mathcal N_q(\alpha)=
\exp\left[-|\alpha|^2(1-1/q^2)/2 \right] /c_q$, and
$|\alpha^*/q\rangle$ is a coherent state in mode $a$ defined by $a
|\alpha^*/q \rangle=(\alpha^*/q) |\alpha^*/q\rangle $. Thus, the
output state of mode $b$ is
\begin{equation}\label{e15}
\rho_{\alpha} = |\mathcal N_q(\alpha)|^2\;{_a\langle \alpha^*/q\;
|\Omega_q \;|\alpha^*/q\rangle_a}\, .
\end{equation}
Let $|Z_a,Z_b\rangle$ be a two-mode coherent state defined by
$(a,b)|Z_a,Z_b\rangle=(Z_a,Z_b)|Z_a,Z_b\rangle$, where $Z_a,Z_b$
are complex amplitudes. Then, the Husimi $Q$-function for
$\Omega_q$ can be defined as
\begin{equation}\label{eq:DefQfun}
 Q_{\Omega_q}(Z_a^*,Z_b^*,Z_a,Z_b)=\langle Z_a,Z_b|\Omega_q |Z_a,Z_b\rangle,
\end{equation}
and the corresponding density operator is the following
normally-ordered operator
\begin{equation}\label{eq:normal1}
\Omega_{q} = :Q_{\Omega_{q}} (a^\dagger,b^\dagger,a,b):\;,
\end{equation}
which is simply the operator functional obtained by replacing the
variables $(Z_a^*,Z_b^*,Z_a,Z_b)$ with $(a^\dagger,b^\dagger,a,b)$
in the $Q$-function given by Eq.~(\ref{eq:DefQfun}), analogously
to Eq.~(\ref{eq:normal2}). Therefore, using Eq.~(\ref{e15}) and
the normally-ordered form of $\Omega_q$, we have the following
simple form for the Husimi $Q$-function
\begin{equation}\label{e16}
 Q_{\rho_\alpha}({Z_b}^*,Z_b) =
|N_q(\alpha)|^2Q_{\Omega_q}({\alpha}/{q},{Z_b}^*,{\alpha^*}/{q},
Z_b )
\end{equation}
of  the output state $\rho_{\alpha}$. Eqs.~(\ref{e15})-(\ref{e16})
are the explicit expressions of the output state for the input  of
{\em any} coherent state $|\alpha\rangle$. According to
Ref.~\cite{Lobino08}, if we know the output states for all input
CSs, then we know the output states of all states in Fock space.
In this approach, given any input state $|\psi\rangle$, we can
write it in its linear superposition form in the coherent-state
basis, and then obtain the $Q$-function of its output state by
using Eq.~(\ref{e16}).

These results can be generalized for a multimode QPT. To apply the
Jamiolkowski isomorphism~\cite{Jamiolkowski72}, we consider $k$
pairs of maximally-entangled states, each in modes $a_1,b_1$,
$a_2,b_2$,$\cdots$, $a_k,b_k$. Explicitly, $ | \Phi^+\rangle =
|\phi^+\rangle_1|\phi^+\rangle_2\cdots|\phi^+\rangle_k. $ Here
$|\phi^+\rangle_i=\lim_{q\rightarrow 1} \exp({qa_i^\dagger
b_i^\dagger})|00\rangle$ indicates a maximally-entangled state in
modes $a_i,\;b_i$.  Subspaces $a$ and $b$ each are now $k$-mode.
Any state $|\psi\rangle$ in subspace $b$, can still be written in
the form of Eq.~(\ref{00}), with the new definitions for
$|\psi\rangle$ and $|\Phi^+\rangle$. Using Eq.~(\ref{phiq}), it is
obvious that the output state of these $k$-pairs of TMSV fully
characterizes the process.

\section{Ancilla-free Gaussian process tomography with a few coherent
states}

Now we present the main result of this paper, which is an
efficient tomography of Gaussian processes probed with only a
\emph{few} single-mode coherent states without the assistance of
ancillas.

As shown in Ref.~\cite{Lobino08}, if we only use CSs in the test,
the tomography of  an unknown process in Fock space requires tests
with {\em all} CSs. Though this problem can be solved by taking
the photon-number-cut-off approximation, in a quantum-optical
process associated with intense light, one still needs a huge
number of different CSs for the test. Here we show that the most
important process in quantum optics, the Gaussian
process~\cite{Eisert05,Holevo12}, can be {\em exactly}
characterized with only a few CSs in the test.

A Gaussian process maps Gaussian states into Gaussian
states~\cite{Eisert02}. Therefore the Husimi $Q$-function of the
operator $\rho_\varepsilon$ must be Gaussian:
\begin{equation}
\label{r0}
Q_{\rho_\varepsilon}(Z_{a}^*,Z_{b}^*,Z_a,Z_{b}) = \exp(c_0+L
+ L^\dagger + S + S^\dagger + S_0 ),
\end{equation}
where
\begin{eqnarray*}
L&=&(\Gamma_a,\Gamma_b)\left(
\begin{array}{c}Z_a\\Z_b\end{array}\right), \\
S&=&\frac{1}{2}(Z_a,Z_b) X
\left(\begin{array}{c}Z_a\\Z_b\end{array}\right),\\
S_0&=&(Z_a^*,Z_b^*)Y\left(\begin{array}{c}Z_a\\Z_b\end{array}\right),
\\
X&=& X^T=\left(\begin{array}{cc}X_{aa} & X_{ab}\\
X_{ba} & X_{bb}\end{array}\right),\\
Y &=&Y^\dagger
=\left(\begin{array}{cc}Y_{aa} & Y_{ab}\\ Y_{ba} & Y_{bb}
\end{array}\right).
\end{eqnarray*}
Before testing the map, all these  are unknowns. The
normally-ordered form of the density operator $\rho_{\varepsilon}$
is
\begin{equation}\label{eq:normal2}
\rho_{\varepsilon}=:Q_{\rho_{\varepsilon}}(a^\dagger,b^\dagger,a,b):
\end{equation}
corresponding to Eq.~(\ref{r0}) but with variables
$(Z_a^*,Z_b^*,Z_a,Z_b)$ replaced by
$(a^\dagger,b^\dagger,a,b)$. The normal order notation
$:\ldots:$ indicates that any term inside it is reordered by
placing the creation operator in the left. For example,
$:aba^\dagger b^\dagger a:=a^\dagger b^\dagger
a^2b$.

The output state from any single-mode input coherent state $|
u\rangle $ (in mode $b$) is
\begin{equation}
\rho_{ u} = {\rm tr}_a\left[ \left(| u^*\rangle\langle
u^*|\right)_a \otimes I \cdot \rho_{\varepsilon}\right],
\end{equation}
where $\rho_{\varepsilon}$ is given by Eq.~(\ref{eq:normal2}). Its
Husimi $Q$-function is
\begin{eqnarray}\label{cohq}
Q_{\rho_{ u}}(Z_b^*,Z_b) &=& Q_{\rho_\varepsilon}( u, Z_b^*,
u^*,Z_b)\nonumber\\
&=& \exp(c_u+L_{ u} + L_{ u}^\dagger + R + R^\dagger + {R_0}
),\hspace{7mm}
\end{eqnarray}
where
\begin{eqnarray*}
  &L_u = (\Gamma_b + u^*X_{ab} + uY_{ab}) Z_b, &\\
  &R = Z_b X_{bb} Z_b/2, \quad  {R_0} = Z_b^* Y_{bb}Z_b,&
\end{eqnarray*}
and $c_u$ is determined by $c_0$, $\Gamma_a$, $X_{aa}$, and
$Y_{aa}$. Explicitly,
\begin{equation}\label{fator}
c_u =  c_0+{\rm Re}\left(2\Gamma_a u^* +  u^* X_{aa} u^* +
uY_{aa}u^*\right).
\end{equation}
The quadratic functional terms ($R,R^\dagger,R_0$) in the exponent
in Eq.~(\ref{cohq}) are independent of $u$; these terms must be
the same for the output states from any input CSs. Therefore,
these can be known by testing the map with one coherent state.
Thus, we do not need to consider these terms below.

Now suppose that we test the process with six different CSs,
$|\alpha_i\rangle$, and $i=1,2,\cdots, 6$. Assume also that
the detected Husimi $Q$-function of the output states is
\begin{equation}\label{qa}
Q_{\rho_{\alpha_i}}(Z^*_b,Z_b)  = \exp(c_i+D_{i} + D_{
i}^\dagger + R + R^\dagger + {R_0} ),
\end{equation}
where $D_i = d_i Z_b$ is the detected (hence known) linear term.
We note that there are available efficient methods of
Gaussian QST based on homodyne detection, which enable the estimation
of the Wigner function or, equivalently, the Husimi $Q$-function
for Gaussian states~\cite{Rehacek09}. According to
Eq.~(\ref{cohq}), the $Q$-function of the output state from the
initial state $|\alpha_i\rangle$ of mode $b$ must be
\begin{equation}
Q_{\rho_{\alpha_i}}(Z^*_b,Z_b) = Q_{\rho_\varepsilon}( \alpha_i,
Z_b^*, {\alpha_i}^*,Z_b).
\end{equation}
Therefore,  we can derive self-consistent equations  by using the
detected data from $\rho_{\alpha_i}$ and setting $u=\alpha_i$ in
Eq.~(\ref{cohq}):
 \begin{equation}\label{self}
  L_i=D_i,\quad c_{\alpha_i} = c_i,
\end{equation}
where $L_i $, $c_{\alpha_i}$ are just $L_u$, $c_u$, respectively,
after setting $u=\alpha_i$ in Eqs.~(\ref{cohq})-(\ref{fator});
$D_i$ and $c_i$ are known from tests. Explicitly,
\begin{equation}\label{eq:Li}
  L_i=(\Gamma_b + {\alpha_i}^*X_{ab}+ {\alpha_i}Y_{ab}) Z_b.
\end{equation}
The first part of Eq.~(\ref{self}) causes:
\begin{eqnarray}\label{KK}
K\cdot
 \left( \Gamma_b,\; X_{ab},\;Y_{ab} \right)^T = d,
\end{eqnarray}
where
\begin{equation*}
K=\left(
\begin{array}{ccc}
1 & {\alpha_1}^* & \alpha_1\\
1 & {\alpha_2}^* & \alpha_2\\
1 & {\alpha_3}^* & \alpha_3
\end{array}
\right) ,\quad
d=\left(\begin{array}{c}
  d_1 \\ d_2 \\ d_3
\end{array}\right).
\end{equation*}
There are three unknowns  ($\Gamma_b$, $X_{ab}$, and
$Y_{ab}$) with three equations now. We find
\begin{equation}\label{root}
\left( \Gamma_b,\; X_{ab},\;Y_{ab} \right)^T = K^{-1}d\;.
\end{equation}

If the Gaussian process is known to be trace-preserving, then
Eq.~(\ref{root})  completes the tomography: up to a numerical
factor, we can deduce all the output states of the other input
CSs, $|\alpha_i\rangle$, for $i=4,5,6$. The term $c_i$ can be
fixed through normalization, which is determined  by the quadratic
and linear functional terms in the exponent of the $Q$-functions.
Knowing these $\{c_i\}$, one can construct $\rho_{\varepsilon}$
completely as shown below.

For any map, $c_i$ can be known from tests with
$|\alpha_i\rangle$, for $i=1,2,\cdots,6$.  We then have
\begin{equation}\label{JJ}
J\cdot \left( c_0,\;  \Gamma_a,\;  \Gamma^*_a,\;  X_{aa},\;
X^*_{aa},\;  Y_{aa} \right)^T=c\;,
\end{equation}
where
\begin{equation*}
J=\left(\begin{array}{cccccc} 1 & \alpha^*_1 & \alpha_1 &
{1\over 2}\alpha^{*2}_1 & {1\over 2}\alpha^2_1 & |\alpha_1|^2
\\ 1 & \alpha^*_2 & \alpha_2 & {1\over 2}\alpha^{*2}_2 &
{1\over 2}\alpha^2_2 & |\alpha_2|^2 \\ 1 & \alpha^*_3 &
\alpha_3 & {1\over 2}\alpha^{*2}_3 & {1\over 2}\alpha^2_3 &
|\alpha_3|^2 \\ 1 & \alpha^*_4 & \alpha_4 & {1\over
2}\alpha^{*2}_4 & {1\over 2}\alpha^2_4 & |\alpha_4|^2 \\ 1 &
\alpha^*_5 & \alpha_5 & {1\over 2}\alpha^{*2}_5 & {1\over
2}\alpha^2_5 & |\alpha_5|^2 \\ 1 & \alpha^*_6 & \alpha_6 &
{1\over 2}\alpha^{*2}_6 & {1\over 2}\alpha^2_6 & |\alpha_6|^2
\end{array}\right), \; c=\left(\begin{array}{c} c_1 \\ c_2\\
c_3\\c_4\\c_5\\c_6
\end{array}\right),
\end{equation*}
for the second part of Eq.~(\ref{self}). Thus
\begin{equation}\label{rootJ}
\left( c_0,\;  \Gamma_a,\;  \Gamma^*_a,\;  X_{aa},\;
X^*_{aa},\; Y_{aa} \right)^T=J^{-1}c\; .
\end{equation}
 {\bf Theorem}: Given $K$ and $J$ defined by
Eqs.~(\ref{KK})-(\ref{JJ}), then the QPT of any single-mode
Gaussian process in Fock space can be performed   with six input
CSs, when $\det K\not=0$ and $\det J\not=0$. The QPT of any
trace-preserving single-mode Gaussian process in Fock space can be
executed  with three input CSs, when $\det K\not=0$.

For example, one can simply choose $\alpha_1=0$,
$\alpha_2=1$, $\alpha_3=i$, $\alpha_4=-1$, $\alpha_5=-i$, and
$\alpha_6=1+i$. One finds
\begin{eqnarray}\label{28}
c_0&=&c_1,\quad \Gamma_b=d_1, \nonumber \\
\Gamma_a&=&\frac14\left(c_2+ic_3-c_4-{i }c_5\right), \nonumber \\
X_{ab}&=&\frac12\left[-{(1+i) }d_1+d_2+{i }d_3\right], \nonumber \\
Y_{ab}&=&\frac12\left[-{(1-i) }d_1+d_2-{i}d_3\right],  \\
Y_{aa}&=&\frac14(c_2+c_3+c_4+c_5)-c_1, \nonumber\\
X_{aa}&=&\frac14\left[c_2-c_3+c_4-c_5+2i(c_1-c_2-c_3+c_6)\right],\nonumber
\end{eqnarray}
where $\{d_i\}$  and $\{c_i\}$ are defined in Eq.~(\ref{qa}).

\subsection{Example: Output state of a beam-splitter process}

As  a check of  our conclusions, we calculate the output
state of a beam-splitter (BS) process as shown in Fig.~1. The
BS has input modes $b$ and $c$ and output modes $b'$ and
$c'$. Regarding this as a black-box process, the only input
is mode $b$ and the only output is mode $b'$. We set mode $c$
to be the vacuum. The BS transforms the creation operators of
modes $b$ and $c$ by:
\begin{equation}\label{change}
U_{\rm BS}\left(b^\dagger,\;  c^\dagger \right)U^{-1}_{\rm
BS}=\left(b^\dagger,\;  c^\dagger \right)M_{\rm BS},
\end{equation}
where $M_{\rm BS}=\left(\begin{array}{cc}\cos\theta
&\sin\theta
 \\ -\sin\theta &\cos\theta
\end{array}\right)$.
If we test such a process with a coherent state
$|\alpha_i\rangle$, we shall find $\rho_{\alpha_i}=
|\alpha_i\cos\theta\rangle\langle\alpha_i\cos\theta|$. Comparing
this with Eq.~(\ref{qa}), we have $d_i=\alpha^*_i\cos\theta$ and
$c_i=-|\alpha_i\cos\theta|^2$. Using
Eqs.~(\ref{root})-(\ref{rootJ}), we find
\begin{eqnarray}
Y_{bb}=-1 ,\; X_{ab}=\cos\theta,\;
Y_{aa}=-\cos^2\theta,\nonumber\\
\Gamma_a=\Gamma_b=Y_{ab}=X_{aa}=X_{bb}=c_0=0.\;
\end{eqnarray}
Therefore
\begin{equation}
\rho_{\varepsilon} =:\exp(a^\dagger b^\dagger \cos\theta
-a^\dagger a \cos^2\theta-b^\dagger b+a b \cos\theta): .
\end{equation}
With this we can predict the output state of {\em any} input
state, for example the squeezed coherent state (squeezed
displaced vacuum)
\begin{equation}
  \ket{\xi(r,Z)}=\exp\left[{{r \over
  2}(b^2}-b^{\dagger 2})\right]\exp\left({Z b^\dagger-Z^* b }\right)\ket{0},
\end{equation}
where $r$ is real. According to  Eq.~(\ref{e12}),
\begin{equation}\label{eq:rhoxi}
  \rho_\xi ={\rm tr}_a\left[(\ket{\xi(r,Z^*)}\bra{\xi(r,Z^*)})_a \otimes I_b \cdot
  \rho_{\varepsilon}\right].
\end{equation}
As a result,
\begin{equation}
Q_{\rho_\xi}(Z^*_b,Z_b)=C\exp(\mathcal H_1 -\mathcal H_2 +
\mathcal H_3 + \mathcal H_4),
\end{equation}
where $C$ is the normalization factor, and
\begin{eqnarray*}
\mathcal H_1 &=& |Z_b|^2(\tanh^2 r\sin^2 \theta-1 )/g , \\
\mathcal H_2 &=& (Z^2_b+{Z_b^*}^2)\tanh r\cos^2\theta /(2g),\\
\mathcal H_3 &=& Z_b\cos\theta (Z^*-Z\tanh r\sin^2\theta)/(g \cosh r),\\
\mathcal H_4 &=& Z^*_b \cos\theta (Z-Z^*\tanh
r\sin^2\theta)/(g \cosh r),
\end{eqnarray*}
and $g=1-\tanh^2 r\sin^4\theta$. This is the same result
obtained from direct calculations using Eq.~(\ref{change}).
\begin{figure}\label{f1}
  \includegraphics[width=150pt]{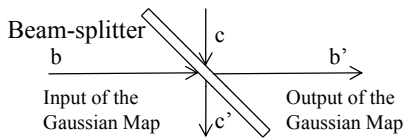}\\
\caption{Gaussian map constructed by a beam-splitter.
Here, we assume that the input mode $c$ is in the
vacuum.}\label{BS}
\end{figure}

\section{Efficient multimode-Gaussian QPT}

Multimode Gaussian QPT has many important applications. For
example, it applies to a complex linear optical circuit  with BSs,
squeezers, homodyne detections, linear losses, Gaussian noises,
and so on. Consider now a Gaussian process acting on a $k$-mode
input state (in modes $b_1,b_2,\cdots, b_k$), with outcome also a
$k$-mode state. Even though other methods~\cite{Lobino08} can also
be extended to the multimode case, the number of input states
required there increases exponentially with the number of modes
$k$, because the number of ket-bra operators
$|\{n_i\}\rangle\langle \{m_i\}|$ in Fock space increases
exponentially with $k$. As shown below, the number of input states
in our method increases {\em polynomially}.

A $k$-mode QPT can be tested with $k$-mode CSs, if the process is
Gaussian. The main Eqs.~(\ref{root})-(\ref{rootJ}) still hold
after redefining the notations there. First,
$\Gamma_a,\;\Gamma_b$, $u$, $\alpha_i$, $d_i$, $Z_a$, and $Z_b$
are now $k$-mode vectors. For example,
\begin{eqnarray*}
  &|\alpha_i\rangle = |\alpha_{i1},\alpha_{i2},\cdots,\alpha_{ik}\rangle,& \\
  &d_i = (d_{i1}, d_{i2},\cdots, d_{ik}), \quad
  Z_b = (Z_{b1},Z_{b2},\cdots, Z_{bk}),&
\end{eqnarray*}
and so on. Following Eq.~(\ref{r0}), $\mathcal {X}_{xy}$ is now a
$k\!\times\! k$ matrix, for $\mathcal X = X$ or $Y$ with $x=a,b$;
$y=a,b$.  We still apply Eqs.~(\ref{root})-(\ref{rootJ}) to
calculate \{$\Gamma_b$, $X_{ab},\; Y_{ab}$\} and \{$\Gamma_a$,
$X_{aa}$, $Y_{aa}$\}, respectively, but keep in  mind that the
matrices $K$, $J$ and symbols $d$, $c$ are now redefined. There
are $(2k +1)k$ unknowns in ($\Gamma_B,\; X_{ab},\; Y_{ab}$). We
need $(2k+1)$ different CSs of $k$-mode to fix these unknowns. Now
we have
\begin{eqnarray}\label{eq:kkm}
K = \left(\begin{array}{ccc}1 & \alpha_1^* & \alpha_1\\
1 & \alpha_2^* & \alpha_2\\
\vdots & \vdots & \vdots \\
 1 & \alpha_{2k+1}^* & \alpha_{2k+1}\end{array}\right),
\end{eqnarray}
which is a $(2k+1)\times (2k+1)$ matrix, since each $\alpha_i$
here is a $k$-mode row vector. Moreover, $d$ is here a
$(2k+1)\times k$ matrix as $d^T= \left(d_1^T,\;d_2^T, \cdots,
d_{2k+1}^T \right)$, with $d_i = (d_{i1},d_{i2}, \cdots, d_{ik})$.
Similarly,
\begin{eqnarray}\label{jjm}
J = \left(\begin{array}{cccccc}1 & \alpha_1^* & \alpha_1 & {1\over 2}\alpha^{*2}_1& {1\over 2}\alpha^2_1 &|\alpha_1|^2\\
1 & \alpha_2^* & \alpha_2 & {1\over 2}\alpha^{*2}_2& {1\over 2}\alpha^2_2 &|\alpha_2|^2\\
\vdots & \vdots & \vdots & \vdots & \vdots & \vdots \\
 1 & \alpha_N^* & \alpha_N & {1\over 2}\alpha^{*2}_N& {1\over 2}\alpha^2_N &|\alpha_N|^2\end{array}\right),
\end{eqnarray}
which is now a $N\times N$ matrix, and $N=(k+1)(2k+1)$, since
$\alpha_i^2$ and $|\alpha_i|^2$ here are row vectors of
$\alpha^2_i=(E_{i1},\;E_{i2},\cdots, E_{ik}) $ and
$|\alpha_i|^2=(\tilde E_{i1},\;\tilde E_{i2},\cdots, \tilde
E_{ik})$, and each element of $E_{im}$ (or $\tilde E_{im}$) is a
vector with ($k-m+1$) modes (or $k$ modes), as
\begin{eqnarray*}
  E_{im} &=& (\alpha_{im}^2,\alpha_{im}\alpha_{i,m+1},\alpha_{im}\alpha_{i,m+2},\cdots, \alpha_{im}\alpha_k,\alpha_k^2), \\
 \tilde E_{im}&=& (\alpha_{im}\alpha^*_{i1},\alpha_{im}\alpha^*_{i2},\cdots, \alpha_{im}\alpha^*_{i,k-1},\alpha_{im}\alpha^*_{ik}).
\end{eqnarray*}
Obviously, $c$ is a column vector with $N$ elements. Therefore we
conclude with this:
\\{\bf Corollary}:
Any $k$-mode Gaussian QPT can be performed with $(k+1)(2k+1)$
different CSs of $k$-mode; or with $(2k+1)$ different CSs of
$k$-mode if the process is trace preserving.

\section{Conclusions}

In summary, we have presented explicit formulas for the tomography
of quantum-optical Gaussian processes probed with only a few
different coherent states. We have reduced the problem of Gaussian
QPT to Gaussian QST, for which efficient methods are
experimentally available~\cite{Rehacek09}. We have extended our
results to multimode Gaussian QPT and demonstrated that the number
of test states required increases only polynomially with the
number of modes.

\acknowledgments

X.B.W. is supported by the National High-Tech Program of China,
Grants No. 2011AA010800, No. 2011AA010803, and No. 2006AA01Z420,
NSFC Grants No. 11174177 and No. 60725416, and the 10 000 Plan of
Shandong province. A.M. acknowledges support from the Polish
National Science Centre under Grant No. DEC-2011/03/B/ST2/01903.
F.N. acknowledges partial support from the ARO, RIKEN iTHES
Project, MURI Center for Dynamic Magneto-Optics, JSPS-RFBR
Contract No.~12-02-92100, Grant-in-Aid for Scientific Research
(S), MEXT Kakenhi on Quantum Cybernetics, and FIRST (Funding
Program for Innovative R\&D on S\&T).


\end{document}